\documentstyle[prl,aps]{revtex}

\begin{document}

\draft

\preprint{\today}

\title{Interplane Transport and Superfluid Density in Layered
Superconductors}
            \author{S.V.~Dordevic$^1$, E.J.~Singley$^1$, D.N.~Basov$^1$,
J.H.~Kim$^{1}$,
M.B.~Maple$^{1}$ and E.~Bucher$^2$}
            \address{$^1$ Department of Physics, University of California,
San Diego, La
Jolla, CA 92093}
            \address{$^2$ Lucent Technologies, Murray Hill, NJ 07974}

\wideabs{
\maketitle

\begin{abstract}
We report on generic trends in the behavior of the interlayer
penetration depth $\lambda_c$ of several  different classes of quasi
two-dimensional superconductors including cuprates, Sr$_2$RuO$_4$,
transition
metal dichalcogenides and organic materials of the $(BEDT-TTF)_2X$ -
series.
Analysis of these trends reveals two distinct patterns in the scaling
between the values of $\lambda_c$ and the magnitude of the DC
conductivity:
one realized in the systems with a Fermi liquid (FL) ground state and the
other seen in systems with a marked deviation from the FL response.
The latter pattern is found primarily in under-doped cuprates and
indicates
a dramatic enhancement (factor $\simeq 10^2$) of the energy scale
$\Omega_C$ associated with the formation of the condensate compared to the
data for the FL materials. We discuss implications of these results for
the
understanding of pairing in high-$T_c$ cuprates.
\end{abstract}
}

\narrowtext

The formation of the superconducting  condensate in elemental metals and
their alloys  is well understood within the theory of Bardeen, Cooper and
Schrieffer (BCS) in terms  of a pairing instability in the ensemble of
Fermi liquid quasiparticles.  Applicability of the FL description to
high-$T_c$ cuprate superconductors is challenged by  remarkable  anomalies
found in both the spin- and charge response of these compounds in the normal
state \cite{orenstein00}. Because  quasiparticles are not well defined at
$T>T_c$ in most cuprates it is natural to inquire into the distinguishing
characteristics of the superconducting condensate which appears to be
built
from entirely different "raw material". Infrared spectroscopy is perfectly
suited for the task. Indeed, the analysis of the optical constants in far
infrared (IR) unfolds the process of the formation of the condensate
$\delta(0)$-peak in the dynamical conductivity and also  gives insight
into
the single-particle excitations in a system both above and below $T_c$
\cite{basov99}.

In this paper we focus on the interplane properties of high-$T_c$
superconductors. We will show that the distinctions in the behavior of the
condensate in the FL superconductors and high-$T_c$ cuprates are most
radical in the case of the $c$-axis interplane response. The analysis of
the generic trends seen in the behavior of the $c$-axis condensate
(correlation between the penetration depth $\lambda_c$ and the DC
conductivity) allows us to infer the energy scale $\Omega_C$ associated
with the development of the superfluid. This energy scale is of the
order of the energy gap $2\Delta$ in FL superconductors but may
dramatically exceed the gap in systems lacking well defined
quasiparticles at $T>T_c$ (primarily in under-doped cuprates).  We
discuss  a connection between the magnitude of $\Omega_C$ and the nature
of the normal state response.

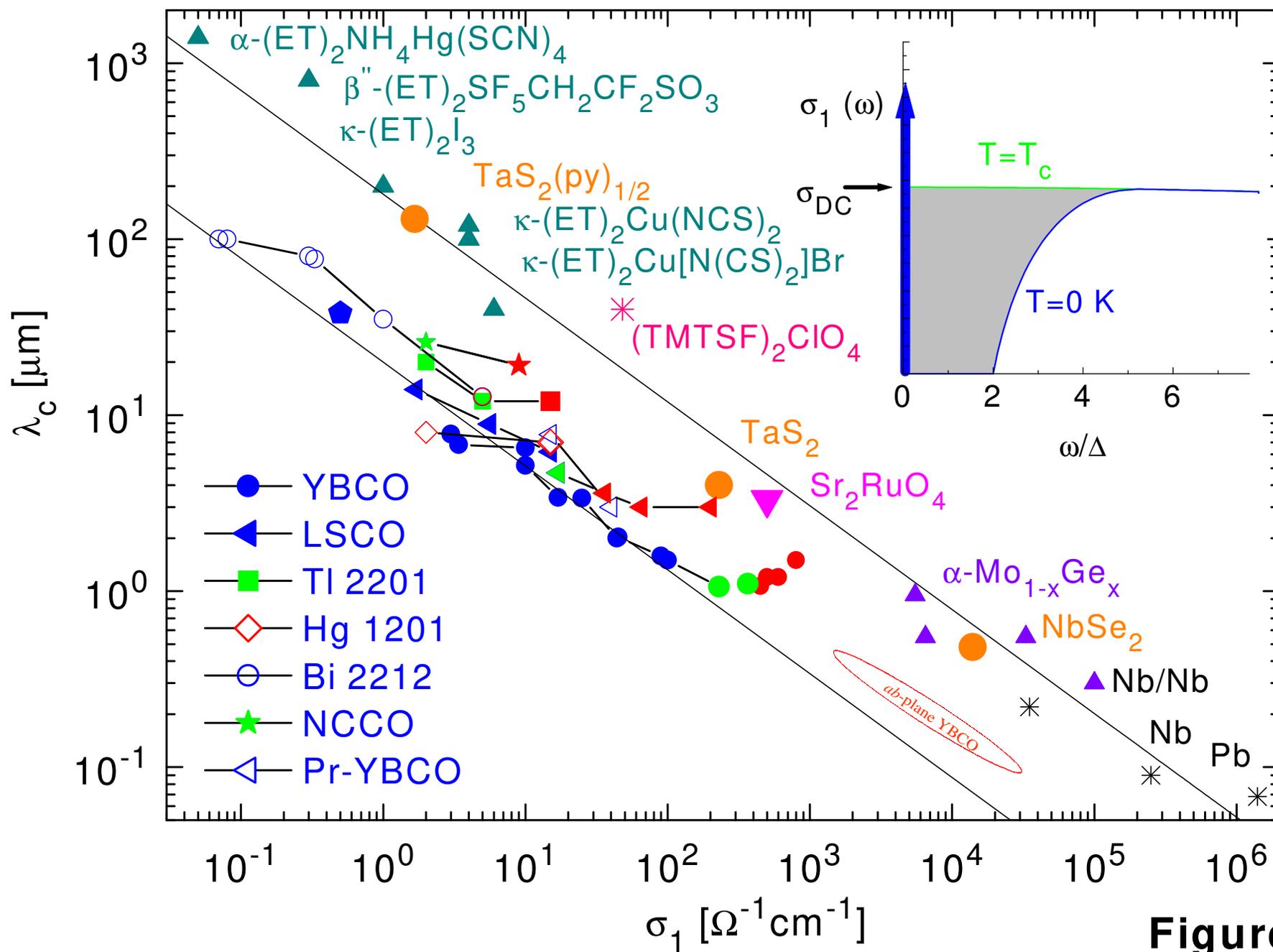
\begin{figure}
\caption{The $c$-axis penetration depth reveals two patterns of
scaling behavior between the magnitude of the interlayer
penetration depth $\lambda_c(T=0K)$ and $dc$-conductivity
$\sigma_{DC}(T_c)$. Most cuprate superconductors exhibit much
shorter penetration depths than non-cuprates materials with the
same $\sigma_{DC}(T_c)$. This result implies dramatic enhancement
of the energy scale $\Omega_C$ from which the condensate is
collected as described in the text. Data points:
YBCO [4,21,22], overdoped YBCO [4,8,23], La214 [5,22,24],
HgBa$_2$Cu$_2$O$_4$ [25], Tl2201 [2,3,9],
Bi$_2$Sr$_2$CaCu$_2$O$_8$ [26], Nd$_{2-x}$Ce$_x$CuO$_4$ [27].
Blue points - underdoped, green - optimally
doped; red - overdope. Transition metal dichalcogenides [13],
(ET)$_2$X compounds [28], (TMTSF)$_2$ClO$_4$ [29], Sr$_2$RuO$_4$
[30], niobium [31,32], lead [32], niobium Josephson junctions [33] and
$\alpha$Mo$_{1-x}$Ge$_{x}$ [34]. Inset: in a
conventional dirty limit superconductor the spectral weight of the
superconducting condensate (given by $1/\lambda^2$) is collected primarily
from the energy gap region. The total normal weight is preset by magnitude
of $\sigma_{DC}$ whereas the product of $\Delta\times \sigma_{DC}$
quantifies the fraction of the weight that condenses.} \label{fig:basov}
\end{figure}

Experimentally, the inter-layer $c$-axis penetration depth of a
layered superconductor can be measured using a variety of
techniques such as microwave absorption, magnetization, vortex
imaging \cite{kam} or IR spectroscopy \cite{ybco124}. In the IR
approach, the magnitude of $\lambda_c$ can be reliably extracted
from the imaginary part of the complex conductivity
$\sigma_1(\omega)+i\sigma_2(\omega_2)$ as $1/\lambda^2 =
\omega\times\sigma_2(\omega)$. Regardless of the method employed,
the penetration depth in several families of cuprates reveals a
universal scaling behavior with the magnitude of
$\sigma_{DC}(T=T_c)$ (Fig.~\ref{fig:basov})\cite{ybco124}: the
absolute value of $\lambda_c$ is systematically suppressed with
the increase of the normal state conductivity. The scaling is
obeyed primarily in under-doped cuprates. The deviations from the
scaling are also systematic and are most prominent in over-doped
phases (red symbols in Fig.~1). Such deviations  are a direct
consequence of a well-established fact: on the over-doped side of
the phase diagram  the DC conductivity increases whereas $\lambda_c$
is either unchanged or may show a minor increase
\cite{uchida96,panagop00}.

We find a similar scaling pattern in other classes of layered
superconductors, including organic materials, transition metal
dichalcogenides and Sr$_2$RuO$_4$ (Fig.~\ref{fig:basov}). While the
non-cuprate data set is not nearly as dense, the key trend is analogous
to the one found for cuprates. The slope of the $\lambda_c-\sigma_{DC}$
dependence
is also close for both cuprates and non-cuprate materials. The principal
difference is that the cuprates universal line is shifted down by
approximately one order of magnitude in $\lambda_c$. The latter result
shows that the superfluid density ($\propto 1/\lambda^2$) is significantly
enhanced in under-doped cuprates compared to non-cuprate materials  with
the same DC conductivity.

Possible origins of the $\lambda_c-\sigma_{DC}$ correlation were recently
discussed in the literature \cite{theory}. A plausible qualitative
account of this effect can be based on the Ferrel-Glover-Tinkham (FGT) sum
rule:
 \begin{equation}
\frac{c^2}{\lambda^2}=\int_{0+}^{\Omega_C}[\sigma_1^N(\omega)-
\sigma_1^{SC}(\omega)]d \omega
\label{eq:eq1}
\end{equation}
where $\sigma^N(\omega)$ is the normal state conductivity and
$\sigma^{SC}(\omega)$ is the
conductivity due to un-paired carriers in the superconducting state at
$\omega>0$ and $T\ll T_c$. For a dirty limit superconductor
$\sigma_1^N(\omega)\approx\sigma_{DC}$, and Eq.~\ref{eq:eq1}
can be approximated as:
 \begin{equation}
\frac{c^2}{\lambda^2}\approx 2\Delta\sigma_{DC}.
\label{eq:eq2}
\end{equation}
Such an approximation is possible because within the BCS model the energy
scale $\Omega_C$ from which the condensate is collected is of the order of
magnitude of the gap: $\Omega_C\simeq 2\Delta\simeq 3-5 kT_c$. A
connection between $1/\lambda^2$, $\sigma_{DC}$, and the energy gap is
illustrated in the inset of Fig.~1. In the dirty limit the magnitude of
$\sigma_{DC}$ sets the amount of spectral weight available in the normal
state conductivity whereas the magnitude of $\Omega_C\simeq 2\Delta$
defines the fraction of this weight which is transferred into condensate
at $T<T_c$. Therefore, the magnitude of $\lambda_c$ can be expected to
systematically decrease with the enhancement of the DC conductivity, in
accord with the FGT sum rule. Notably, an approximate form
(Eq.~\ref{eq:eq2}) yields the $\lambda_c(\sigma_{DC})$ scaling with the
power law $\alpha=1/2$ which is close to $\alpha = 0.59$ seen in
Fig.~\ref{fig:basov}.

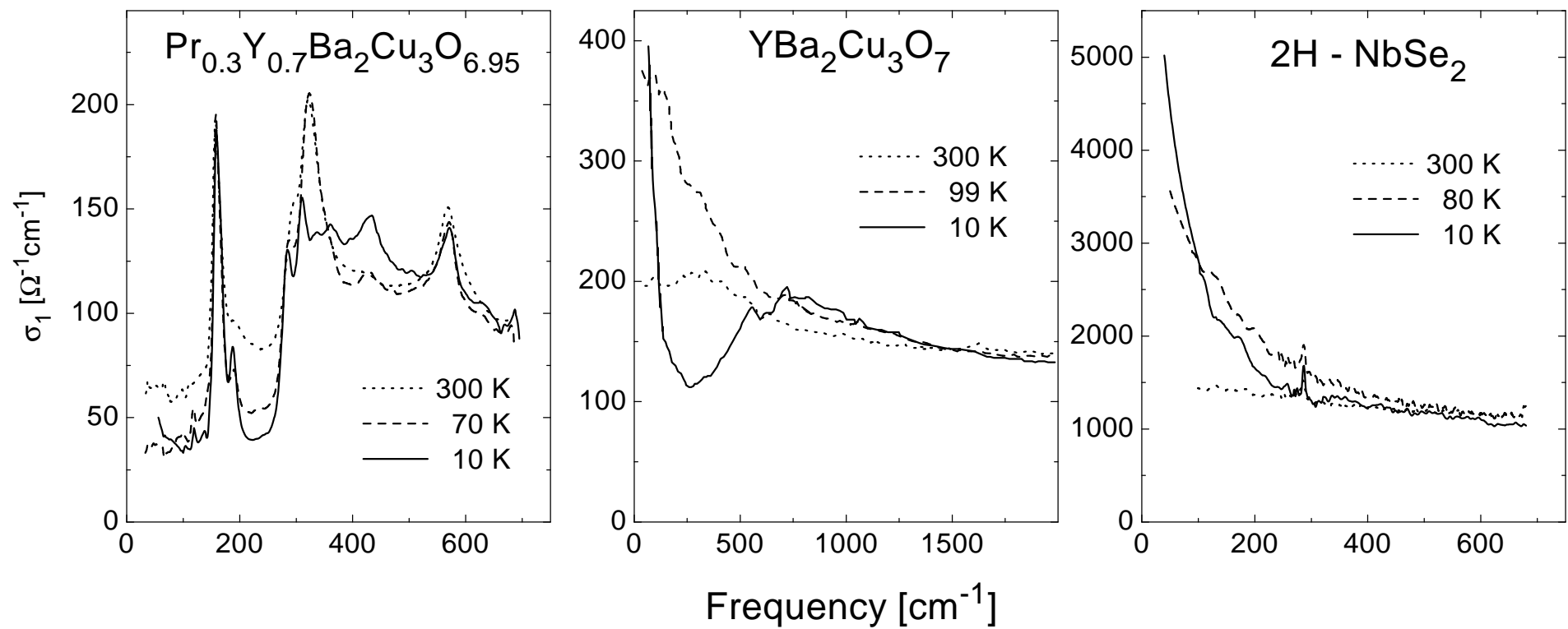
\begin{figure}
\caption{Examples of the interplane optical conductivity for layered
superconductors. The observation of the Drude feature in the interplane
optical conductivity of the dichalcogenide 2H-NbSe$_2$ (right panel) is
consistent with magnetoresistance measurements that revealed evidence for
well-behaved quasiparticles. Instead, the conductivity of the underdoped
Pr$_{0.3}$Y$_{0.7}$Ba$_2$Cu$_3$O$_{6.95}$ material (left panel) suggests
no
signs of coherent response.
Overdoped cuprates show the emergence of the Drude feature (middle panel)
and also occupy intermediate position between non-FL and FL data set in
Fig.~\ref{fig:basov}.}
\label{fig:cond}
\end{figure}

Interestingly, the sum rule arguments discussed above successfully
describe the universal $\lambda_c - \sigma_{DC}$ scaling in the underdoped
cuprates despite the fact that the superconducting energy gap is not
well-defined in the interlayer conductivity of these materials. The gap-less
response  of cuprates is exemplified in left panel in Fig.~\ref{fig:cond}
displaying $\sigma_1(\omega)$ data for
Pr$_{0.3}$Y$_{0.7}$Ba$_2$Cu$_3$O$_{6.95}$ with $T_c\simeq 60$ K. The
conductivity was determined through a Kramers-Kronig analysis of
reflectivity measured over the frequency range from 10 cm$^{-1}$ to 48,000
cm$^{-1}$. Regardless of the low- and high-energy extrapolations, the
$\sigma_1(\omega)$ data is unchanged over the energy interval displayed in
Fig.~\ref{fig:cond}. The response of the Pr-doped sample is similar to the
features found in the conductivity of under-doped oxygen deficient
YBa$_2$Cu$_3$O$_{7-\delta}$ (YBCO) \cite{bernhardpr}. Namely, the normal
state conductivity is suppressed as the sample is cooled down to $T_c$,
with a transfer of spectral weight to higher energies. Below $T_c$, one
does not find any radical changes in the $\sigma_1(\omega)$ spectra of
Pr-doped YBCO or in the oxygen-deficient YBCO with a similar $T_c$. Most
importantly, this system, along with all other under-doped compounds,
shows significant absorption in the superconducting state so that
$\sigma_1^{SC} > 0$. Therefore, in cuprates only a {\it small fraction} of
the far-infrared spectral weight is contributing to the
condensate. The latter result is in apparent conflict with the assumption
$\sigma_1^{SC}(\omega<2\Delta)\simeq 0$, which allows one to reduce
Eq.~\ref{eq:eq1} to an approximate form given by Eq.~\ref{eq:eq2}.
However,
the strong condensate density seen in the cuprates located on the
universal
line can be understood in terms of the dramatic enhancement of the energy
scale $\Omega_C$ over  the magnitude  of the energy gap. This latter
conclusion
also follows from the explicit sum rule analysis for samples of underdoped
La$_{2-x}$Sr$_x$CuO$_4$ (La214) and YBCO materials, suggesting that
$\Omega_C$ in these compounds exceeds 0.1-0.2 eV \cite{basov99}.

The involvement of a broad energy scale into the formation of
the condensate in underdoped cuprates is also supported through a
comparison of the  universal scaling patterns seen for these
materials and of a similar pattern detected for non-cuprate
superconductors. According to Eqs.~\ref{eq:eq1} and \ref{eq:eq2},
the energy scale associated with the condensate formation for
materials on the upper line is of the order of 3 meV which is
close to estimates of the gap for most "conventional" materials in
Fig.~1. If Eq.~\ref{eq:eq2} is employed to describe the difference
between the upper and the lower lines in Fig.~\ref{fig:basov},
than one is forced to conclude that the corresponding scale for
underdoped cuprates is of the order of $\simeq$ 0.3 eV. As pointed
out above, this assessment of $\Omega_C$ is supported by the
explicit sum rule analysis\cite{basov99} and also makes $\Omega_C$
the largest energy scale in the problem of cuprate
superconductivity. Data points in Fig.~\ref{fig:basov} for
overdoped materials support the notion that the
$\lambda_c-\sigma_{DC}$ plot provides means to learn about the
energy scale associated with the condensate formation. Deflection
of the  over-doped cuprates from the universal line implies that
$\Omega_C$ is gradually suppressed with the increased carrier
density. This trend is common for Tl$_2$Ba$_2$CuO$_{6+\delta}$
(Tl2201), La214 and YBCO materials (see Fig.~\ref{fig:basov}).
Integration of the conductivity for all these materials shows that
the FGT sum rule is exhausted at energies as low as 0.08 eV
\cite{katz00,basov00}.

A quick inspection of the materials in Fig.~1 suggests that the $\simeq 3$
meV scale is observed in systems in which superconductivity emerges out of
a FL state whereas the enhanced value of $\Omega_C\simeq$ 0.3 eV is found
in non-FL underdoped cuprates. The experiments which in our opinion are
most relevant to FL versus non-FL classification include quantum
oscillations of the low-$T$ inter-layer resistivity (and of other
quantities) in high magnetic fields \cite{amro}. Quantum oscillations is a
direct testimony of long-lived quasiparticles which are capable of
propagating coherently between the layers. Such quasiparticles are a
prerequisite  of the Fermi-liquid ground state. On the contrary, quantum
oscillations have never been reported for under-doped cuprates. The lack
of
coherence in the $c$-axis transport in these materials indicates that the
ground state of cuprates is distinct from the FL picture.

Signatures of FL versus non-FL behavior also can be recognized in the
spectra of the $c$-axis conductivity. A hallmark of the FL response is
the Drude peak seen in $\sigma_1(\omega)$ of metals. Notably, this
feature is never found in underdoped compounds (forming the non-FL line in
Fig.~\ref{fig:basov}). The electronic contribution to $\sigma_1(\omega)$
in
these materials is usually structureless which is commonly associated with
the incoherent (diffusive) motion of charge carriers across the planes.
On the contrary, many materials that belong to the upper (FL) line in
Fig.~\ref{fig:basov} demonstrate a familiar Drude behavior. This
behavior has been found in Sr$_2$RuO$_4$ \cite{katsufuji} and is also
shown
in our  data for inter-plane response of NbSe$_2$ (Fig.~\ref{fig:cond},
right panel), reported here for the first time. In both cases, the width
of
the peak decreases at low temperatures, which is characteristic of the
response of ordinary metals. Earlier measurements of the $c$-axis thermo
reflectance proved applicability of conventional BCS electrodynamics
to the NbSe$_2$ data\cite{nbse2}. As for the over-doped
compounds (located in a cross-over region between non-FL and FL lines in
Fig.~\ref{fig:basov}) their conductivity is indicative of the formation of
the Drude peak (Fig.~\ref{fig:cond}, middle panel), which is becoming more
pronounced with the increase of the carrier density.

To summarize the experimental results, we wish to stress the following
points: {\it i)} two distinct patterns in $\lambda_c-\sigma_{DC}$
correlation (Fig.~\ref{fig:basov}) originate from a dramatic difference
($\simeq 10^2$) in the energy scale $\Omega_C$ from which the interlayer
condensate is collected; {\it ii)} the pattern with the typical energy
scale of $\simeq 3$ meV is realized in the materials with the coherent FL-
type transport between the planes, whereas the one with $\Omega\simeq
300$ meV is found in cuprate superconductors with the incoherent non-FL
response; {\it iii)} over-doped cuprates reveal a cross-over between the
two behaviors. These results allow us to draw several conclusions
regarding
the features of the superconducting condensate in different
layered systems\cite{pimenov}.  \\
 $\bullet$ The symmetry of the order
parameter seems to be unrelated to trends seen in the condensate response.
Indeed, the upper line in Fig.~\ref{fig:basov} is formed by  $s$-wave
transition metal dichalcogenides, $p$-wave Sr$_2$RuO$_4$ and organic
materials for which both $s$- and $d$-wave states have been proposed
\cite{amro}, while $d$-wave high-$T_c$ materials form the lower line and
the crossover region between the lines. \\
 $\bullet$  Electrodynamics of the FL systems at $T\ll T_c$ is determined
by the magnitude of the gap (and hence by $T_c$), in general
agreement with the BCS theory. It is therefore hardly surprising
that the trend initiated by layered FL materials is also followed
in 1-dimensional organic conductors as well as  by more
conventional systems such as Nb Josephson junctions, bulk Nb, Pb or
amorphous $\alpha$Mo$_{1-x}$Ge$_x$ (see Fig.~\ref{fig:basov}).\\
 $\bullet$ We find no obvious connection between the broad energy scale
$\Omega_C$ and the critical temperature of the studied superconductors.
While scaling of $\lambda_c$ by the magnitude of $T_c$ does
reduce the "scattering" of the data points, the two distinct $\lambda_c-
\sigma_{DC}$  patterns  persist even if such scaling is implemented. In
particular, the critical temperature of under-doped La214 materials is
nearly the same as that of the several ET-compounds ($\simeq 12-15$
K). Nevertheless, the penetration depth is dramatically enhanced in the
latter system. \\
 $\bullet$ While the pseudogap state was shown to be responsible for
the anomalous superfluid response of the underdoped
cuprates\cite{basov00}, the characteristic pseudogap temperature $T^*=90-
350$ K is still much lower than our estimate of $\Omega_C$ for these
materials.

Dramatic enhancement of the $\Omega_C$ scale well beyond the gap value in
underdoped cuprates may be connected with a  well-established feature of
the electronic spectral function $A(\omega)$ in underdoped cuprates. At
$T>T_c$ the spectral function shows no quasiparticle peak in the antinodal
direction; the spectral weight of $A(\omega)$ appears primarily in the
incoherent channel and is spread out over the energy interval extending
beyond 0.3-0.5 eV\cite{arpes}. The coherent contribution to $A(\omega)$ is
observed  in the superconducting state and the $T$-dependence of its
weight is similar to that of the superfluid density.\cite{arpes2} These
experiments suggest that the emergence of the condensate is associated
with the changes in the incoherent region of the spectral function. We stress a
quantitative agreement between our estimate of the $\Omega_C$ and the
energy region affected at $T<T_c$ in the ARPES measurements.\cite{arpes3}
Because carrier condensation is connected with the perturbation of the
entire spectral function, the  scale set by the superconducting gap
becomes
irrelevant to the problem of the interplane electrodynamics of underdoped
cuprates in accord with our findings. The  development of coherence in the
spectral function at $T<T_c$ may be consistent with the lowering of the
electronic kinetic energy \cite{norman}, a result which was inferred
earlier from the sum rule analysis of the interlayer conductivity
\cite{basov99}.

Close correspondence between the properties of the spectral
function in the $(\pi,0)$ region and the features of the
interlayer conductivity may stem from the peculiarities of the
interlayer transfer matrix elements which are maximized for this
particular region \cite{matrix}. Therefore it is natural to look
for a common microscopic origin of the anomalies in the $c$-axis
response and characteristic features of the ARPES spectra at
$(\pi,0)$. The latter has been associated with the electron
fractinalization at $T>T_c$ \cite{orgad}. An important attribute
of  scenarios of cuprate superconductivity discussed in Ref.\cite{orgad}
is that at $T<T_c$ the total energy is lowered due to reduction of
the kinetic energy, in qualitative agreement with the data.
A quantitative account of the distinct energy scales associated
with the condensate in non-FL materials is a challenge for models
attempting to describe superconductivity in the cuprates. This research was
supported by the U.S. DOE, NSF and Research Corporation.

\end{document}